\documentclass[aps,twocolumn,showpacs,preprintnumbers,amsmath,amssymb]{revtex4}
\usepackage{epsfig}

\newcommand{\beq}{\begin{eqnarray}}
\newcommand{\eeq}{\end{eqnarray}}
\newcommand{\diff}{\mbox{d}}
\newcommand{\pardis}{\langle \mu \rangle}

\newcommand{\ie}{{\it i.e.\ }}

\begin{document}     

  
\title{Magnetic charge superselection in the deconfined phase of 
Yang-Mills theory}
 
\author{M. D'Elia$^{a,1}$, A. Di Giacomo$^{b,2}$, B. Lucini$^{c,3}$}

\affiliation{$^a$  Dipartimento di Fisica dell'Universit\`a di Genova and INFN,
Sezione di Genova, Via Dodecaneso 33, I-16146 Genova, Italy}
\affiliation{$^b$ Dipartimento di Fisica dell'Universit\`a di Pisa
and INFN, Sezione di Pisa, Via Buonarroti 2 Ed. C, I-56127 Pisa, Italy}
\affiliation{$^c$ Theoretical Physics, University of Oxford,
1 Keble Road, OX1 3NP Oxford, UK}
\affiliation{$^1$ e-mail address: delia@ge.infn.it}
\affiliation{$^2$ e-mail address: digiaco@df.unipi.it}
\affiliation{$^3$ e-mail address: lucini@thphys.ox.ac.uk}
\affiliation{$$}
\affiliation{$$}

\begin{abstract}
The vacuum expectation value of an operator carrying magnetic charge is
studied numerically for temperatures
above the deconfinement temperature in SU(2) and SU(3)
gauge theory. By analyzing its finite size behaviour, this is found to
be exactly
zero in the thermodynamical limit for any $T > T_c$ whenever the magnetic
charge of the operator is different from zero. These results show that magnetic
charge is superselected in the hot phase of quenched QCD.
\end{abstract}

\maketitle

\section{Introduction}              
\label{sectintroduction}
Color confinement is one of the most elusive phenomena in QCD. A natural
working hypothesis is that confinement is the consequence of the breaking of
some symmetry. A possible candidate is the symmetry related to Abelian
magnetic charge conservation \cite{thooft81}. The underlying idea 
is that a description of the theory exists in terms of dual fields, which have
non-zero magnetic charge and are weakly coupled in the confined phase.
This fits the idea of dual superconductivity of the vacuum
as the mechanism of color confinement \cite{mandelstam,thooft76}.

With this scenario in mind, an operator $\mu$ carrying magnetic charge
has been constructed in terms of the fundamental variables of the theory
\cite{zacopane,oldartu1,artu1,oldartsu2} for several Abelian projections and
its vacuum expectation value $\pardis$
has been studied across the deconfinement phase transition in SU(2)
\cite{artsu2} and SU(3) \cite{artsu3} gauge theories by lattice calculations.
It has been shown that the magnetic symmetry is
broken in the confined phase and implemented {\em \`a la Wigner} in the
deconfined phase. For the quenched theory the confined and the deconfined phase
are defined in terms of the string tension or in terms of the vacuum
expectation value of the Polyakov line.
 In the critical region below the deconfinement temperature $T_c$,
$\pardis$ scales with the critical indices of the transition. This shows the
interconnection between confinement-deconfinement and breaking-restoring of
the magnetic symmetry. $\pardis$ is indeed a disorder parameter for the
deconfining phase transition. This analysis has been extended
to full QCD with similar results \cite{artqcd1,artqcd2} and an extension
to SU($N$) is on the way \cite{artsun}. In the full QCD case the Polyakov
line is not an order parameter and the transition is indicated by a drop
of the chiral condensate, which is not either an order parameter at non-zero
quark mass, so that $\pardis$ is the only available (dis)order parameter and
could provide a genuine criterion for confinement.

It has been found numerically \cite{artsu2,artsu3,artran} and then
supported by analytical arguments \cite{digiacomo} that the behaviour 
of $\pardis$ is
independent of the Abelian projection, implying that dual
superconductivity is an intrinsic property of the confining vacuum.

While $\pardis$ has been studied in great detail in the confined 
(low temperature) region up to the critical temperature $T_c$, 
several aspects have still to be explored regarding the behaviour
of $\pardis$ in the deconfined (high temperature) region. 
In this regime the magnetic symmetry is implemented 
{\em \`a la Wigner}, so that the Hilbert space of the theory
is split into mutually orthogonal sectors labelled by the value of the
magnetic charge ({\em superselection}). As a consequence, in the thermodynamic
limit $\pardis = 0$ exactly for any $T > T_c$ if $\mu$ carries a net magnetic 
charge that is different from zero. While it has already been 
found~\cite{artsu2,artsu3,artran} that, in the case where $\mu$ creates 
a single magnetic monopole, $\pardis$ goes to zero exponentially as a
function of the lattice size (\ie in the thermodynamical limit) 
in the extreme weak coupling (infinite temperature) limit, a detailed 
finite size study of $\pardis$ in the whole deconfined region down to $T_c$
is still lacking. Such a study would reveal whether $\pardis$
is indeed strictly zero for any $T > T_c$. Moreover, 
a more extensive study performed on operators creating various 
monopole configurations with either zero or non-zero net magnetic charge
would clarify if the observed phenomenon is really
a manifestation of magnetic charge superselection in the 
deconfined phase of Yang-Mills theories.

It is the purpose of the present work to give an answer to those open 
questions. We do that both for SU(2) and SU(3) pure gauge theories.

The rest of the paper is organised as follows. In Sect.~\ref{sectgen}
we give the technical details of our lattice simulations.
In Sect.~\ref{sectsu3} we discuss our results
for SU(3) and SU(2) gauge theory. Finally, Sect.~\ref{sectconclusions}
summarizes our conclusions.
\section{The Calculation}
\label{sectgen}
On the lattice, the disorder parameter is defined as
\beq
\label{eq:mu}
\langle \mu \rangle =
\frac {\int \left( {\cal D} U \right) e^{- \beta S_M}}
{\int \left( {\cal D} U \right) e^{- \beta S}} \ ,
\eeq
where $S$ is the Wilson action and $S_M$ is the ``monopole''
action, obtained from $S$ by modifying the temporal plaquette at time zero
by the procedure developed in Refs.~\cite{artsu2,artsu3,artu1}.
By this procedure a single monopole can be created or a generic 
configuration of monopoles and antimonopoles.

Note that, because of the
abelian projection independence of $\pardis$, 
there is no need to perform explicitely
an Abelian projection as done in Refs.~\cite{artsu2,artsu3,artsun}. To use the
terminology introduced in \cite{artran}, we will be studying $\pardis$ in
a ``random projection''.

In order to study the dependence on the 
total magnetic charge, we have computed $\pardis$  in the extreme 
weak coupling for monopoles of different charges and for dipole
field configurations obtained by putting two monopoles with opposite
magnetic charge at distance $d$.
We have also investigated ``double monopole'' configurations obtained
by putting two monopoles of equal charge at distance $d$.

If confinement is related to magnetic charge symmetry as
discussed above, we expect that, in the deconfined region, 
$\pardis = 0 $ in the infinite volume limit
whenever the net magnetic charge is non-zero and that it can stay
different from zero if the net magnetic charge is zero.

Instead of $\pardis$, we computed numerically
\beq
\label{eq:rho}
\rho = \frac{\diff}{\diff \beta} \log \pardis = \langle S \rangle_S - 
\langle S_M \rangle _{S_M} \ ,
\eeq
where the subscript indicates the action with respect to which the average 
has been taken.
$\rho$ carries the same physical information as $\pardis$ and is much easier
to determine in Monte Carlo simulations \cite{oldartu1,oldartsu2,artu1}.
The value of $\pardis$ is related to $\rho$ by the relationship
\beq
\pardis = \exp\left(\int_0^{\beta} \rho(\beta^{\prime})\mbox{d}\beta^{\prime}\right) \; .
\label{reconstr}
\eeq
As it is clear from Eq.~\ref{reconstr}, $\lim_{N_s \to \infty} \pardis = 0$
in the whole deconfined region ($T > T_c$, $\beta > \beta_c$) if
$\lim_{N_s \to \infty} \rho = - \infty$ for every $\beta > \beta_c$. 

In our investigation, we used lattices of sizes $N_s^3 \times 4$ with $N_s$
ranging from 16 to 48. Measurements have been performed sampling over
5000-10000 configurations. The statistical errors are shown in the figures.
Other technical aspects of our simulations are similar to those described in
\cite{artsu2,artsu3}.

In the following section we shall discuss in details our numerical
results for SU(3) and SU(2).
\section {Numerical results}
\label{sectsu3}

\begin{figure}[tbhp]
\vspace{0.5cm}
\centerline{\epsfig{figure=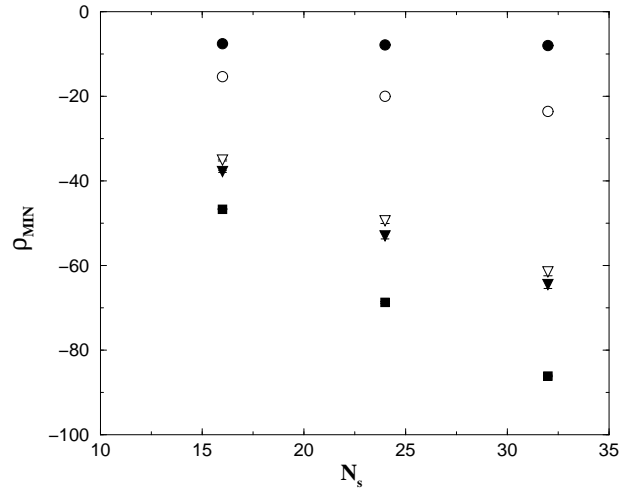,angle=0,width=82mm,height=66mm}}
\vskip 0.86cm
\caption{$\rho$ parameter in the weak coupling limit 
as a function of the spatial lattice size for different values of
the whole magnetic charge in SU(3) gauge theory.
Open circles refer to a single monopole of charge 2, filled circles refer to a
dipole of zero net magnetic charge, filled triangles refer to a single monopole
of charge 4, open triangles refer to two monopoles of charge 2 put at a
distance of 2 lattice spacings apart from each other,
filled squares refer to two monopoles of opposite
charge but different kind ($\lambda_3$ and $\lambda_3'$ generators)
put at a distance of 2 lattice spacings  apart from each other.
}
\label{fig1}
\end{figure}     

\begin{figure}[tbhp]
\centerline{\epsfig{figure=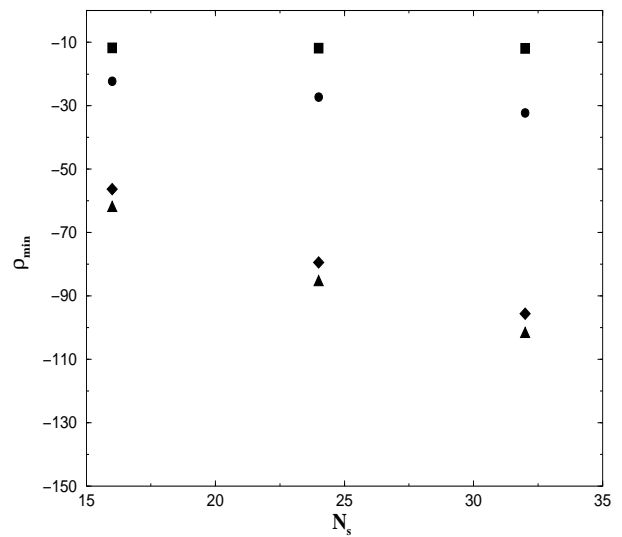,angle=0,width=80mm,height=70mm}}
\vskip 0.5cm
\caption{$\rho$ parameter in the weak coupling limit 
as a function of the spatial lattice size for different values of
the whole magnetic charge in SU(2) gauge theory.
Bullets refer to a single monopole
of charge 2, squares refer to a dipole of zero net magnetic charge,
diamonds refer to a single monopole of charge 4, triangles 
refer to two monopoles of charge 2 put at a distance of 2 lattice spacings
apart from each other.}
\label{figweak}
\end{figure}     

\begin{figure}[tbhp]
\centerline{\epsfig{figure=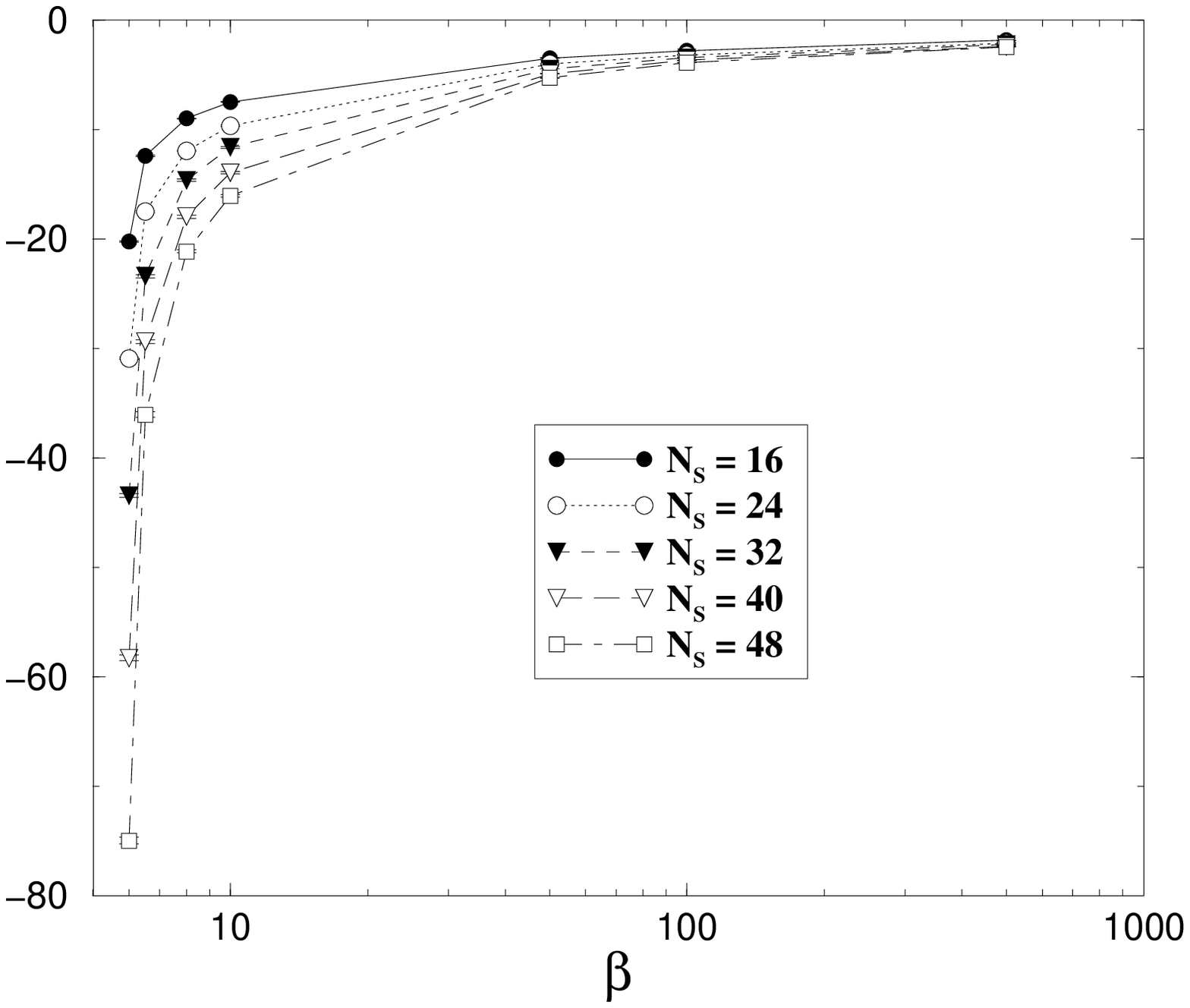,angle=0,width=80mm,height=70mm}}
\vskip 0.5cm
\caption{$\rho/N_s$ as a function of $\beta$ for various lattice
sizes in the deconfined region for a single monopole of charge 2.
SU(3) gauge theory.} 
\label{fig2}
\end{figure}     

\begin{figure}[tbhp]
\centerline{\epsfig{figure=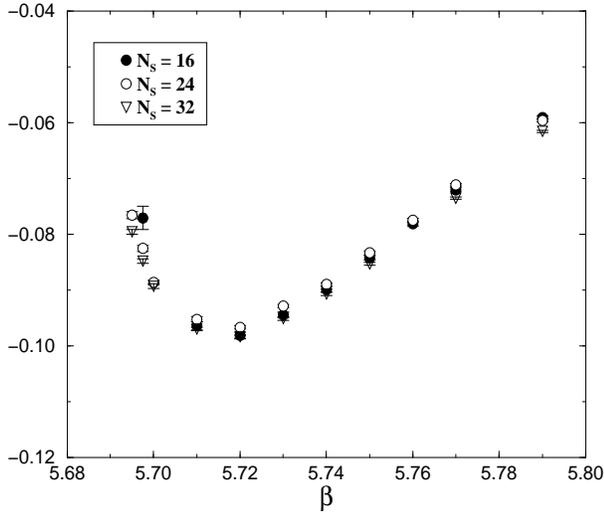,angle=0,width=80mm,height=70mm}}
\vskip 0.5cm
\caption{$\rho/N_s^3 + 1.6/N_s$ in the deconfined region slightly above
$T_c$. SU(3) gauge theory.}
\label{fig3}
\end{figure}     
\vskip0.3cm
\begin{figure}[tbhp]
\vspace{0.0cm}
\centerline{\epsfig{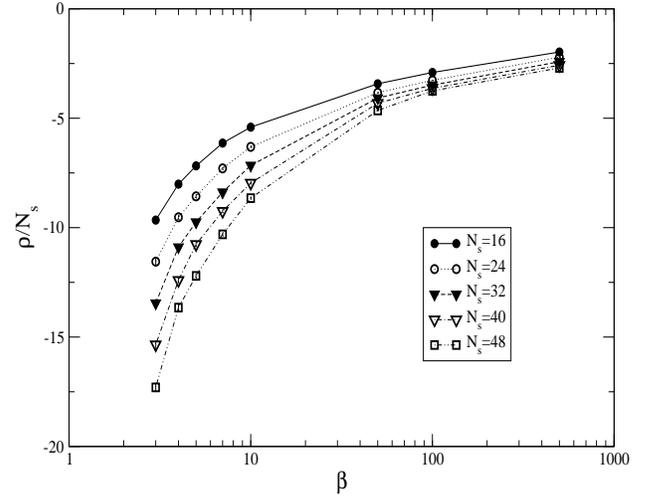}}
\vskip 0.0cm
\caption{$\rho/N_s$ as a function of $\beta$ for various lattice
sizes in the deconfined region for a single monopole of charge 2.
SU(2) gauge theory.}
\vskip0.8cm
\label{figrhoweaksu2}
\end{figure}

\begin{figure}[tbhp]
\centerline{\epsfig{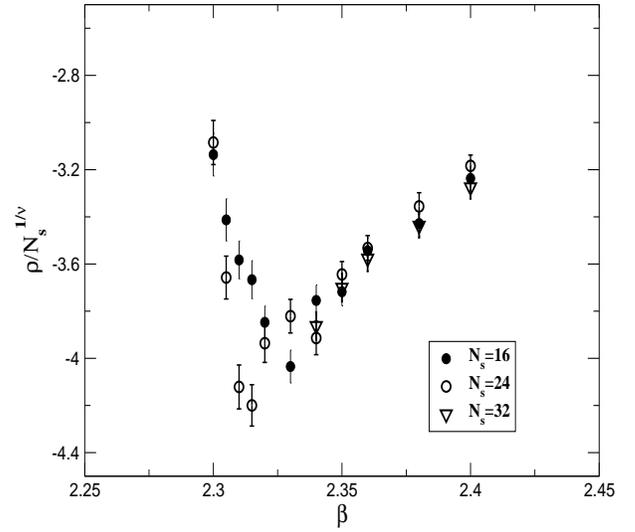}}
\vskip 0.5cm
\caption{Finite size behaviour of $\rho$ in the deconfined region slightly
above $T _c$. SU(2) gauge theory.}
\label{fig1strongc}
\end{figure}

Let us first discuss the results obtained in the extreme weak coupling
limit for different operators $\mu$ creating different monopole
configurations.
In Fig.~\ref{fig1} and in Fig.~\ref{figweak} we show,
respectively for SU(3) and SU(2) pure gauge theories, results obtained 
for $\rho$ as a function
of $N_s$ in the limit $\beta \to \infty$ for different monopole configurations.
As already explained in Refs.~\cite{artsu2,artsu3}, in order to perform
this limit one has to find the absolute minima taken over the ensemble of 
gauge configurations
for both $S$ and $S_M$: the minimum of $S$ is trivially obtained
on the configuration with all the links equal to one, while for $S_M$
we have applied a simulated annealing procedure till a very
high value of $\beta$ followed by straigth cooling protracted till
the value of the minimum stabilizes.

As it is clearly visible from the figures,
$\rho(\beta = \infty)$ diverges linearly with $N_s$ whenever
the net magnetic charge is different from zero and it reaches
a constant value as a function of $N_s$ otherwise\footnote{
One has to be careful in studying the infinite volume limit of $\rho$. The 
overall signal is obtained by integrating, over larger and larger volumes
as $N_s$ increases, a signal density which 
fades out as the distance from the monopole (or dipole) center increases:
so it happens that the finite precision used in the simulation can artificially
wash out an otherwise linear behaviour with $N_s$, and one as
to increase the precision to recover the correct result. This has been taken
care of by performing the computation on large lattices with increased
precision (single to double).}. It is interesting
to notice that the two cases of a single monopole of charge 4 
and of 2 monopoles of charge 2 at distance $d$ from each other lead
to the same $\rho$ apart from a $N_s$-independent correction which
is most likely due to the interaction between the two monopoles.

We have also studied the finite size behaviour of $\rho$ 
in the whole deconfined region for the case of a single monopole of 
magnetic charge 2. Let us first discuss the case of SU(3).
In Fig.~\ref{fig2} we show the behaviour of $\rho/N_s$ as a function
of $\beta$ for different values of $N_s$: it is clear that, while $\rho$ 
diverges linearly with $N_s$ in the limit $\beta \to \infty$, it diverges
more and more rapidly as $\beta \to \beta_c$ from above
($\beta_c \simeq 5.69$ for $N_t = 4$). In particular we have verified
that, very close to $\beta_c$, $\rho$ scales approximately as $N_s^3$,
plus small corrections proportional to lower powers of $N_s$:
this is apparent from Fig.~\ref{fig3}, where the values of
$\rho/N_s^3 + 1.6/N_s$ obtained for different values of $N_s$ fall exactly
on top of each other when plotted as a function of $\beta$. This is consistent
with the behaviour of $\rho$ on the other side of the phase
transition, where it also diverges as $N_s^{1/\nu}$, with $\nu = 1/3$ (first
order phase transition) for SU(3). 

The finite size behaviour of $\rho$ for SU(2) is shown in
Fig.~\ref{figrhoweaksu2}. 
Again for $\beta \gtrsim \beta_c \simeq 2.29$
the scaling is well described by the
ansatz $\rho \simeq c N_s^{1/\nu}$ (see Fig.~\ref{fig1strongc}), where
$\nu \simeq 0.63$, \ie the value for the 3d Ising model \cite{artsu2}.

\section{Conclusions}
\label{sectconclusions}
We have investigated the finite size scaling behaviour of 
$\rho = \frac{\diff}{\diff \beta} \log \pardis$ in the 
deconfined region of both SU(2) and SU(3) pure gauge theories. 
We have shown that, whenever the magnetic charge created
by $\mu$ is different from zero, $\rho$ diverges to $- \infty$ for
$T > T_c$ as the spatial volume is increased, \ie that $\pardis = 0$
in the thermodynamical limit. This supports the expectation
that in the deconfined phase
of Yang-Mills gauge theories magnetic charge is superselected.
In the case where $\mu$ creates a magnetic monopole, we have shown
that $\rho$ diverges linearly in the weak coupling limit, and
more and more rapidly as $T \to T_c$ from above, where it diverges as
$N_s^{1/\nu}$, thus proving
that $\pardis$ is strictly zero for every temperature $T > T_c$.
The results presented here complete the argument of
\cite{artsu2,artsu3,artran} that $\pardis$ is a disorder parameter
for the deconfinement phase transition.

\section*{Acknowledgements}
This work has been partially supported by MIUR and by EU contract 
No. FMRX-CT97-0122. BL is supported by the Marie Curie Fellowship 
No. HPMF-CT-2001-01131.

\end{document}